\documentclass[pra,aps,amsmath,twocolumn,floatfix]{revtex4}

\usepackage{graphics}

\usepackage{epsfig}

\begin{document}


\title{New experimental and theoretical approach to the 3d $^2$D-level lifetimes of $^{40}$Ca$^+$ }

\author{A.~Kreuter}
\affiliation{Institut f\"ur Experimentalphysik, Universit\"at
Innsbruck, Technikerstra{\ss}e 25, A-6020 Innsbruck, Austria}

\author{C.~Becher}
\email{Christoph.Becher@uibk.ac.at}
\affiliation{Institut f\"ur
Experimentalphysik, Universit\"at Innsbruck, Technikerstra{\ss}e
25, A-6020 Innsbruck, Austria}

\author{G.P.T.~Lancaster}
\affiliation{Institut f\"ur Experimentalphysik, Universit\"at
Innsbruck, Technikerstra{\ss}e 25, A-6020 Innsbruck, Austria}

\author{A.B.~Mundt}
\affiliation{Institut f\"ur Experimentalphysik, Universit\"at
Innsbruck, Technikerstra{\ss}e 25, A-6020 Innsbruck, Austria}

\author{C.~Russo}
\affiliation{Institut f\"ur Experimentalphysik, Universit\"at Innsbruck,
Technikerstra{\ss}e 25, A-6020 Innsbruck, Austria}

\author{H.~H\"affner}
\affiliation{Institut f\"ur Experimentalphysik, Universit\"at
Innsbruck, Technikerstra{\ss}e 25, A-6020 Innsbruck, Austria}

\author{C.~Roos}
\affiliation{Institut f\"ur Experimentalphysik, Universit\"at
Innsbruck, Technikerstra{\ss}e 25, A-6020 Innsbruck, Austria}

\author{W.~H\"ansel}
\affiliation{Institut f\"ur Experimentalphysik, Universit\"at
Innsbruck, Technikerstra{\ss}e 25, A-6020 Innsbruck, Austria}

\author{F.~Schmidt-Kaler}
\affiliation{Institut f\"ur Experimentalphysik, Universit\"at
Innsbruck, Technikerstra{\ss}e 25, A-6020 Innsbruck, Austria}

\author{R.~Blatt}
\altaffiliation{also with: Institut f\"ur Quantenoptik und
Quanteninformation, \"Osterreichische Akademie der Wissenschaften,
Technikerstra{\ss}e 25, A-6020 Innsbruck, Austria}
\affiliation{Institut f\"ur Experimentalphysik, Universit\"at
Innsbruck, Technikerstra{\ss}e 25, A-6020 Innsbruck, Austria}

\author{M.S.~Safronova}
\affiliation{Department of Physics and Astronomy, University of
Delaware, Newark, Delaware 19716, USA}

\date{\today}

\begin{abstract}
We report measurements of the lifetimes of the 3d $^2$D$_{5/2}$ and 3d
$^2$D$_{3/2}$ metastable states of a single laser-cooled $^{40}$Ca$^+$ ion in a
linear Paul trap. We introduce a new measurement technique based on
high-efficiency quantum state detection after coherent excitation to the
D$_{5/2}$ state or incoherent shelving in the D$_{3/2}$ state, and subsequent
free, unperturbed spontaneous decay. The result for the natural lifetime of the
D$_{5/2}$ state of 1168(9)~ms agrees excellently with the most precise
published value. The lifetime of the D$_{3/2}$ state is measured with a single
ion for the first time and yields 1176(11)~ms which improves the statistical
uncertainty of previous results by a factor of four. We compare these
experimental lifetimes to high-precision ab initio all order calculations and
find a very good agreement. These calculations represent an excellent test of
high-precision atomic theory and will serve as a benchmark for the study of
parity nonconservation in Ba$^+$ which has similar atomic structure.
\end{abstract}


\maketitle


\section{Introduction}

The lifetime of the metastable D-levels in Ca$^+$ is of high
relevance in various experimental fields such as optical frequency
standards, 
quantum information and astronomy. 
Trapped ion optical frequency standards \cite{Gill03} and optical clocks
\cite{Diddams01} are based on narrow absorption lines in single laser-cooled
ions. With transition linewidths in the 1~Hz range \cite{Rafac00}, Q-values
(frequency of the absorption divided by its spectral width) of $\approx
10^{15}$ can be achieved. As the lifetimes of the D-levels in Ca$^+$ are on the
order of 1~s, yielding sub-Hz natural linewidths of the D-S quadrupole
transitions, Ca$^+$ has long been proposed as a promising candidate for a
trapped ion frequency standard \cite{Werth89}. Such long lifetimes together
with the ability to completely control the motional and electronic degrees of
freedom of a trapped ion \cite{Leibfried03} make it ideally suited for storing
and processing quantum information \cite{SasuraReview}. In Ca$^{+}$ a quantum
bit (qubit) of information can be encoded within the coherent superposition of
the S$_{1/2}$ ground state and the metastable D$_{5/2}$ excited state
\cite{NaegerlQubit} with very long coherence times
\cite{FerdiCoherence,RoosBell}. In astronomy, absorption lines of Ca$^+$ ions
are used to explore the kinematics and structure of interstellar gas clouds
\cite{Hobbs88,Welty96} and the D-level lifetimes are required for
interpretation of the spectroscopic data.
On the other hand, in theoretical atomic physics Ca$^{+}$ is an excellent
benchmark problem for atomic structure calculations owing to large higher-order
correlation corrections and its similarity to Ba$^+$. The size and distribution
of the correlation corrections  make it ideal for the study of the accuracy of
various implementations of the all-order method.
 The properties of Ba$^+$ are of interest due to studies of
parity nonconservation in heavy atoms and corresponding atomic-physics tests of
the Standard model of the electroweak interaction \cite{Koerber03}.

\begin{figure}[!htb]
\includegraphics[width=7cm]{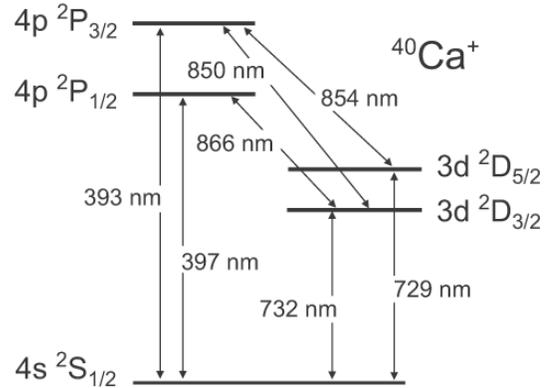}\caption{
Ca$^+$-level scheme with relevant transitions.}\label{levelscheme}
\end{figure}

Experimental investigations of long atomic lifetimes have profited enormously
from the development of ion trap technology and laser spectroscopy. Early
experiments on the measurement of the D-level lifetimes in $^{40}$Ca$^+$
\cite{Urabe92,Arbes93,Arbes94,KnoopPRA95,Gudjons96} used large clouds of ions
and the lifetime was determined by recovery of fluorescence on the
UV-transitions (S$_{1/2}$ - P$_{1/2}$ or S$_{1/2}$ - P$_{3/2}$, see
Fig.~\ref{levelscheme}) after electron shelving in the D-states or by observing
UV fluorescence after driving transitions from the D-states to the P-states.
Shelving in this context means that the electron for a certain time remains in
a metastable atomic level which is not part of a driven fluorescence cycle.
These lifetime measurements were limited by deshelving induced by collisions
with other ions or the buffer gas used for cooling. Similar results using the
same techniques have been obtained in an ion storage ring \cite{Lidberg99}.

Much more accurate results can be obtained by performing lifetime measurements
with single trapped ions \cite{Urabe93,Ritter97,Block99,Barton00,KnoopArxiv} or
strings of few trapped ions \cite{Staanum04} and employing the quantum jump
technique. This technique is based on monitoring the fluorescence on the
S$_{1/2}$ - P$_{1/2}$ dipole transition while at random times the ion is
shelved to the metastable D$_{5/2}$-state where the fluorescence falls to the
background level. For observing fluorescence both the S$_{1/2}$ - P$_{1/2}$
(397~nm) and the D$_{3/2}$ - P$_{1/2}$ (866~nm) transition have to be driven to
prevent the ion from residing in the metastable D$_{3/2}$-state. Shelving to
the D$_{5/2}$-state is initiated by applying laser light at 850~nm (D$_{3/2}$ -
P$_{3/2}$) \cite{Urabe93} or at 729~nm (S$_{1/2}$ - D$_{5/2}$)
\cite{KnoopArxiv}. Statistical analysis of the fluorescence dark times yields
the lifetime $\tau$. The most precise measurement using this technique was
carried out by Barton et al. \cite{Barton00} who found the result of
$\tau$=1168(7)~ms.

Here, we introduce a measurement technique \cite{Kreuter04} based on
deterministic coherent excitation to the D$_{5/2}$ state or incoherent shelving
in the D$_{3/2}$ state, followed by a waiting period with free spontaneous
decay and finally a measurement of the remaining excitation by high-efficiency
quantum state detection. During the waiting time all lasers are shut off and no
light interacts with the ion. This method basically is an improved version of a
technique that was used earlier to measure the D$_{3/2}$ metastable level
lifetime in single Ba$^+$ ions \cite{Yu97}. The main advantage of this "state
detection" method is that no residual light is present during the measurement
which could affect the free decay of the atom. In addition, it allows for the
measurement of the D$_{3/2}$ level lifetime which otherwise is inaccessible
with the quantum jump technique. There exist only a few reported
D$_{3/2}$-level lifetime results for Calcium
\cite{Arbes94,KnoopPRA95,Lidberg99} but none from a single ion experiment.

\begin{figure}[htb]
\includegraphics[width=8cm]{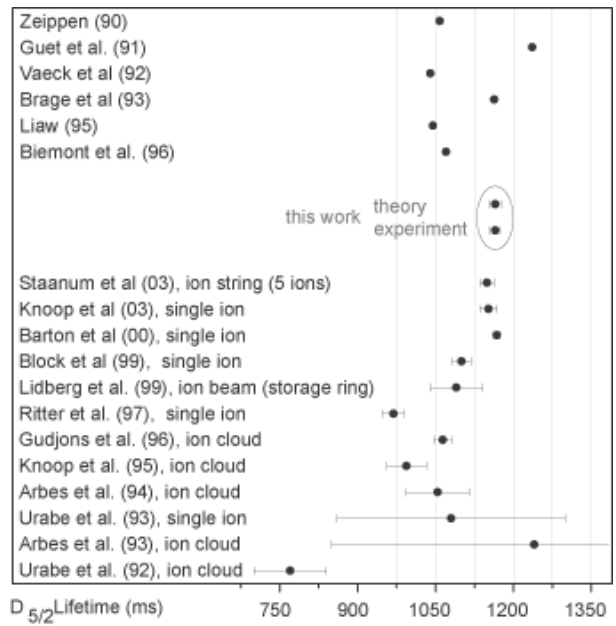} \caption{Theoretical
and experimental results for the D$_{5/2}$-level
lifetime.}\label{D52allresults}
\end{figure}

\begin{figure}[htb]
\includegraphics[width=8cm]{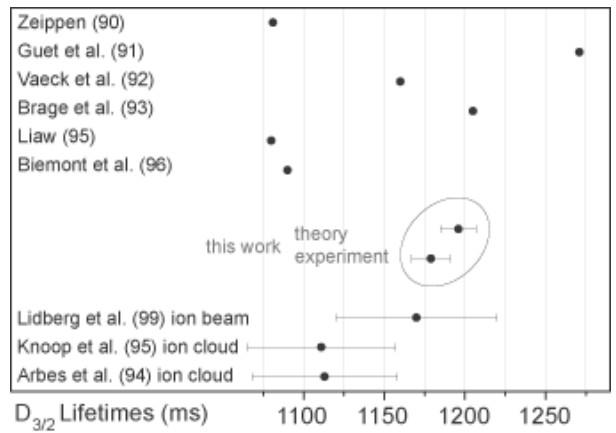} \caption{Theoretical
and experimental results for the D$_{3/2}$-level
lifetime.}\label{D32allresults}
\end{figure}

Figs.~\ref{D52allresults} and \ref{D32allresults} compare the different
experimental
\cite{Urabe92,Arbes93,Urabe93,Arbes94,KnoopPRA95,Gudjons96,Ritter97,Lidberg99,Block99,Barton00,KnoopArxiv,Staanum04}
and theoretical \cite{Zeippen90,Guet91,Vaeck92,Brage93,Liaw95,Biemont96}
methods and results for the D$_{5/2}$- and D$_{3/2}$-level lifetimes.
From Fig.~\ref{D52allresults} it is evident that the single ion measurements
are the most accurate ones. Generally, lifetime measurements on single ions or
crystallized ion strings are more accurate as systematic errors, e.g. due to
collisions, can be reduced to the highest possible extend. Therefore, single
ion D-level lifetime measurements for Calcium are of special interest. The
existence of accurate D-state lifetime values is of special interest for theory
as well since most studies of alkali-metal atoms were focused on the
measurements of the lowest $n$P-state lifetimes and D-states are much less
studied. The properties of D-states are also generally more complicated to
accurately calculate owing to large correlation corrections.


\section{Experimental Setup and Methods}

For the experiments, a single $^{40}$Ca$^+$ ion is stored in a linear Paul trap
in an ultra high vacuum (UHV) environment ($10^{-11}$ mbar range). The Paul
trap is designed with four blades separated by 2~mm for radial confinement and
two tips separated by 5~mm for axial confinement. Under typical operating
conditions we observe radial and axial motional frequencies
$(\omega_{\text{r1}},\omega_{\text{r2}},\omega_{\text{ax}}) =2\pi (4.9, 4.9,
1.7)$~MHz. $^{40}$Ca$^+$ ions are loaded into the trap using a 2-step
photoionization procedure \cite{GuldePhoto}. The trapped $^{40}$Ca$^+$ ion has
a single valence electron and no hyperfine structure (see
Fig.~\ref{levelscheme}). Doppler-cooling on the $\textrm{S}_{1/2} -
\textrm{P}_{1/2}$ transition at 397~nm puts the ion in the Lamb-Dicke regime
\cite{Leibfried03,SasuraReview}. Diode lasers at 866~nm and 854~nm prevent
optical pumping into the D states during cooling and state preparation. For
coherent excitation to the D$_{5/2}$ state we drive the S$_{1/2}$ to D$_{5/2}$
quadrupole transition at 729~nm. A constant magnetic field of 3~G splits the 10
Zeeman components of the S$_{1/2}$ -- D$_{5/2}$ multiplet. We detect whether a
transition to D$_{5/2}$ has occurred by applying the laser beams at 397~nm and
866~nm and monitoring the fluorescence of the ion on a photomultiplier (PMT),
i.e. using the electron shelving technique \cite{shelving}. The internal state
of the ion is discriminated with an efficiency close to 100$\%$ within
approximately 3~ms \cite{Roos99}. The following stabilized laser sources are
used in the experiment: two frequency-stabilized diode lasers at 866~nm and
854~nm with linewidths of $\approx$~10~kHz and two Ti:Sa lasers at 729~nm
($<100$~Hz linewidth) and 794~nm ($<$~100~kHz linewidth), of which the 794~nm
laser is externally frequency doubled to obtain 397~nm. The experimental setup
and the laser sources are described in more detail elsewhere
\cite{FerdiCZReloaded,NaegerlQubit}.


\section{Measurement of the D$_{5/2}$ state lifetime}
\label{secD52lifetime}

\subsection{Measurement procedure and results}

\begin{figure}[htb]
\includegraphics[width=8cm]{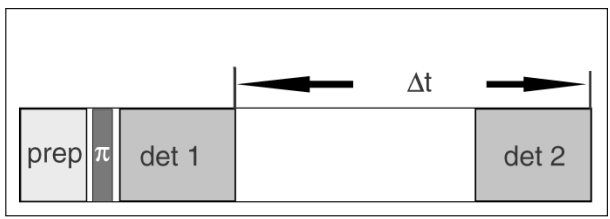}\caption{
Simplified pulse scheme for the D$_{5/2}$ lifetime measurement:
the preparation consists of Doppler cooling, repumping, and
optical pumping (2ms); followed by a $\pi$-pulse (few $\mu$s) and
a detection periods (3.5~ms). The waiting time is varied between
25~ms and 5~s.}\label{D52lt-pulse-simple}
\end{figure}

The measurements consist of a repetition of laser pulse sequences
applied to the ion. The sequence generally is composed of three
steps (see Fig.~\ref{D52lt-pulse-simple}):

1. State preparation and Doppler cooling, consisting of 2~ms of Doppler cooling
(397~nm and 866~nm light), repumping from the D$_{5/2}$ level (854~nm light)
and optical pumping into the S$_{1/2} ($m=-1/2$)$ Zeeman sublevel (397~nm
$\sigma^-$ polarized light).

2. Coherent excitation at 729~nm with pulse length and intensity chosen to
obtain near unity excitation ($\pi$-pulse) to the D$_{5/2} (m=-5/2)$ Zeeman
level.

3. State detection for 3.5 ms by recording the fluorescence on the S$_{1/2}$ -
P$_{1/2}$ transition with a photomultiplier. Discrimination between S and D
state is achieved by comparing the fluorescence count rate with a threshold
value. The state is measured before and after a fixed waiting period $\Delta t$
to determine whether a decay of the excited state has happened.

This three-step cycle is repeated typically several thousand times. The decay
probability $p$ is then determined as the ratio of D-state results in the
second and the first state detections. For the calculation of the
D$_{5/2}$-state lifetime $\tau_{(5/2)}$ we use an exponential fit function
$(1-p) = \exp \{ - \Delta t / \tau_{(5/2)} \}$.
For $\Delta t$ we use the time interval between the ends of the
two detection periods.

\begin{figure}[htb]
\includegraphics[width=8cm]{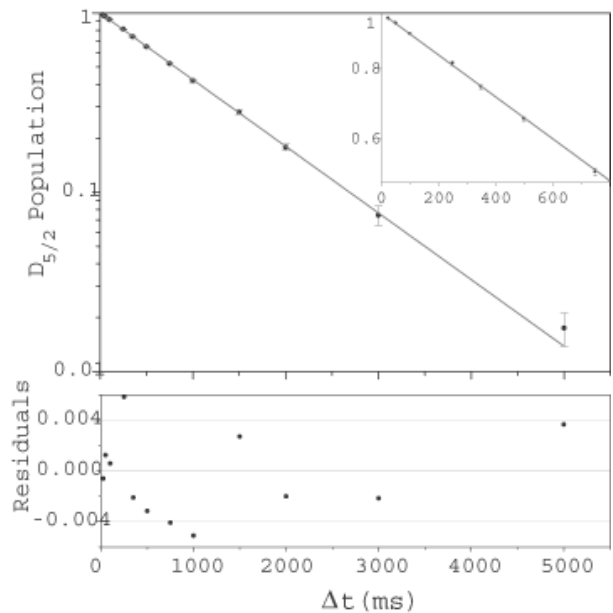}
\caption{D$_{5/2}$-level excitation probability after delay times from 25~ms to
5~s plotted on a logarithmic scale. The solid line is a least squares fit to
the data yielding $\tau_{\textrm{(5/2)}} = 1168(9)$~ms. The residuals
(difference of data points and fit curve) of the fit are shown in the lower
diagram.}\label{D52lifetime}
\end{figure}

Fig.~\ref{D52lifetime} shows the measured D$_{5/2}$-level
excitation probability $(1-p)$ after several delay times ranging
from 25~ms up to 5~s. A weighted least squares fit to the data
yields the lifetime $\tau_{(5/2)} = 1168(9)$~ms 
using the fitting function described above, where the only fitting parameter is
$\tau_{(5/2)}$. The statistical error (in brackets) is the 1$\sigma$ standard
deviation.
The fit yields $\chi^2_{\nu}= 0.47$, indicating that the experimental decay is
consistent with the expected exponential decay behavior. The least-squares
method is justified by the normal distribution of the mean decay probabilities
which is a result of the 'central limit theorem' of statistics. This was also
verified using simulated data sets (see next section).

\subsection{Systematic errors}\label{syserr52}

There are several types of systematic errors that may affect the lifetime
result. In UHV single ion experiments the biggest error source is usually
radiation which irradiates the ion due to insufficient shielding of room light
or insufficient shut-off of laser beams. In our experiment, the strongest
influence stems from residual light at 854~nm. The influence of this radiation
on the D$_{5/2}$ -level lifetime has been investigated extensively in
\cite{Barton00}. If radiation at 854~nm is present during the delay interval it
may de-excite the D$_{5/2}$-level to the ground state via the strongly coupled
P$_{3/2}$-level. This additional "decay channel" artificially shortens the
observed lifetime. The obvious source for residual 854~nm radiation is the
854~nm diode laser itself. In our experiment, it is eliminated by a fast
mechanical shutter \cite{Densitron} which is closed during the delay interval.
The 40~dB attenuation of the double-pass AOM which usually switches the 854~nm
light was shown to be insufficient: In an earlier experiment without the
shutter the lifetime was determined to 1011(6)~ms \cite{FerdiCoherence}. In
addition, the results observed without shutter were found to fluctuate by
approximately $\pm 50$~ms depending on the specific AOM and diode laser
adjustments.

Another source for 854~nm radiation is background fluorescence at 854~nm from
the 866~nm diode laser. To eliminate this radiation an AOM in single pass with
an attenuation of more than 20~dB was used to shut the 866~nm beam. As the
systematic lifetime error without AOM was found to be of the order of a few
percent, this attenuation is sufficient. Note that this source of error cannot,
in principle, be directly eliminated in the quantum-jump technique where 866~nm
light must be radiated onto the ion continuously. In that case, the only way to
correct for this systematic error is to measure at different light powers and
extrapolate the lifetime to zero power which in turn implies a larger
statistical error. In summary, radiation at 854~nm did not influence the
measured D$_{5/2}$ -level lifetime at the given level of statistical
uncertainty.

The D$_{5/2}$ -level lifetime could in principle be also reduced
by transitions between the D-levels, i.e. by a M1-transition
stimulated by thermal radiation. The corresponding transition rate
is given by $W_{12} = B_{12} \rho(\nu)$ with the Einstein
coefficient for stimulated emission $B_{12}$ and the energy
density per unit frequency interval for thermal radiation
$\rho(\nu)$. With the rate of spontaneous emission $A_{12} = (8\pi
h \nu^3/c^3) B_{12}$, $W_{12}$ is rewritten as:
\begin{equation}\label{M1}
    W_{12} = \frac{A_{12}}{\textrm{e}^{h\nu/kT}-1}
\end{equation}
With $\nu=1.82$~THz and $A_{12}=2.45\times10^{-6}$~s$^{-1}$ taken from
\cite{Ali88} we get $W_{12}=7.23\times10^{-6}$~s$^{-1}$ at room temperature
which changes the D$_{5/2}$ -level lifetime by much less than the statistical
error of our measurement.

Residual radiation could also induce lifetime-enhancing systematic effects.
Both radiation at 393~nm (roomlight) or 729~nm (Ti:Sa laser, double-pass AOM
attenuation of $\approx 40$~dB) can lead to re-shelving of the ion. This
effect, however, leads also to a different decay function. It is modeled by a
simple rate equation for the excited state population $p_D$
\begin{equation}\label{rateequation}
\dot{p_D}=-\Gamma p_D + R(1-p_D)
\end{equation}
where $\Gamma = 1/\tau$ denotes the natural decay rate and $R$ the
reshelving rate induced by laser radiation. The solution of
eq.~(\ref{rateequation}) is of the form
\begin{equation}\label{pD}
p_D= A \textrm{e}^{-\Gamma' t} + B
\end{equation}
with $A=1-R/\Gamma'$, an offset $B=R/\Gamma'$ and the new decay rate
$\Gamma'=\Gamma +R$. Thus, an offset $B$ is the signature of a re-shelving
rate. The result from fitting the experimental data in Fig.~\ref{D52lifetime}
with the modified exponential fit function (\ref{pD}) is
$\tau_{\textrm{fit}}=1165(10)$~ms and $R_{\textrm{fit}}=3(2)\times
10^{-3}$~s$^{-1}$.

To evaluate the systematic error due to a repumping rate $R$ we use the
following technique: We generate simulated data sets from random numbers by
considering the fact that the decay probability for a given waiting time is
distributed binomially around a mean that is given by an exponential function
with an expected mean decay time $\tau_{(5/2)}$. For these data sets we also
take into account the particular experimental waiting times and number of
measurements. In this way 'ideal' data sets are created that contain a purely
statistical variation of data according to the experimental settings and that
are free of any systematic errors. First, a fit of Eq.~(\ref{pD}) to such an
ideal simulated data set yields an additional repumping rate $R=0$ with a
standard deviation of $\Delta R=3\times 10^{-3}$~s$^{-1}$. Thus, the above
fitted repumping rate $R_{\textrm{fit}}$ is consistent with zero and not
sufficiently significant to allow any conclusion about the actual rate or the
model, i.e. the statistical error is too large for this small systematic error
to be resolved in a fit to the data. Second, to obtain an upper limit for the
systematic error of the lifetime due to a possible re-pumping rate we assume
that such a rate $R_{\textrm{sim}}$ exists in the experiment. We then simulate
data sets including the rate $R_{\textrm{sim}}$ and the lifetime $\tau_{(5/2)}$
and fit these data sets with a normal exponential fit function $(1-p) = \exp{(
-\Delta t /\tau_{\textrm{sim}})}$. The deviation of the fit result
$\tau_{\textrm{sim}}$ from $\tau_{(5/2)}$ used for the simulation gives the
systematic lifetime error. For $R_{\textrm{sim}}= 3\times 10^{-3}$~s$^{-1}$
this systematic error is $\Delta \tau = -3$~ms.

Another systematic effect that implies a different fit model is the state
detection error. Even though the detection efficiency is close to unity,
Poissonian noise in the PMT counts and the possibility of a spontaneous decay
during the detection period produce a small error \cite{RoosPhD}. The first
error, i.e. the probability $\varepsilon_1$ to measure the ion in the wrong
state due to noise of the count rate, depends on the discrimination between S
and D state in the electron shelving technique. Such discrimination is achieved
by comparing the fluorescence count rate during the detection window with a
threshold value. Proper choice of this threshold value leads to an error as
small as $\varepsilon_1=10^{-5}$ which can be neglected for the following
analysis. The second error, i.e. the probability $\varepsilon_2$ for a wrong
state measurement due to spontaneous decay, also depends on the length of the
detection window, fluorescent count rates for the ion in S and D states and the
threshold setting. For the chosen parameters in the experiment we evaluate
$\varepsilon_2=10^{-3}$. The measured excitation probability is then
$(1-p)_{\textrm{meas}} = (1-\varepsilon_2)(1- p)_{\textrm{real}}$ and implies a
model
function of the form $(1-p)=(1-\varepsilon_2)e^{-\Gamma \Delta t}$. 
A fit to simulated data as described above yields a statistically consistent
limit for this detection error of $\varepsilon_2 = 1\times 10^{-3}$. Again, it
cannot be resolved by a fit to the measured data. From simulated data including
an assumed detection error of $\varepsilon_2 = 1\times 10^{-3}$ we get an upper
limit of the systematic lifetime error $\Delta \tau = 7$~ms.

In addition to radiative effects, non-radiative lifetime reduction mechanisms
exist, namely, inelastic collisions with neutral atoms or molecules from the
background gas. Two relevant types of collisions can be distinguished:
Quenching and j-mixing collisions. Quenching collisions cause direct deshelving
of the ion into the ground state. In the presence of high quenching rates
lifetime measurements have to be done at different pressures. An extrapolation
to zero pressure then yields the natural lifetime. Measurements of collisional
deshelving rates for different atomic and molecular species have been performed
in early experiments \cite{Arbes93,KnoopPRA95}. Ref.~\cite{KnoopPRA95} finds
specific quenching rates for Ca$^+$ of $\Gamma_H^q = 37 \times
10^{-12}$~cm$^3$s$^{-1}$ for H$_2$, and $\Gamma_N^q =170 \times
10^{-12}$~cm$^3$s$^{-1}$ for N$_2$. Collisions may also induce a change of the
atomic polarization, a process called j-mixing or fine structure mixing where a
transition from the D$_{5/2}$ to the D$_{3/2}$ state or vice versa is induced.
These rates have also been determined in Ref.\cite{KnoopPRA95} to $\Gamma^j_H
=3\times 10^{-10}$~cm$^3$s$^{-1}$ for H$_2$ and $\Gamma^j_N =13\times
10^{-10}$~cm$^3$s$^{-1}$ for N$_2$. Such collisional effects cannot be
distinguished from a natural decay process. Collisional effects are most
prominent in experiments with large clouds of ions or at higher background
pressure. To give an upper limit of the effect in our experiment estimates of
the constituents of the background gas must be made. If a background gas
composition of 50\% N$_2$ and 50\% H$_2$ is assumed \cite{Aarhus} and the
pressure $p<2\times10^{-11}$~mbar in the linear ion trap set up is taken into
account, an upper limit for the additional collision induced rate of $\gamma =
3\times 10^{-4}$~s$^{-1}$ is calculated. This effect is well below a $10^{-3}$
relative error and can be neglected here.

In summary, the result for the lifetime of the $D_{5/2}$ level can be quoted
as: $\tau_{(5/2)}= 1168$~ms $\pm 9$~ms~(statistical) -3~ms~(repumping rate)
+7~ms~(detection).



\section{Measurement of the D$_{3/2}$ state lifetime}

\subsection{Measurement procedure and results}

\begin{figure}[!htb]
\begin{center}
\includegraphics[width=8cm]{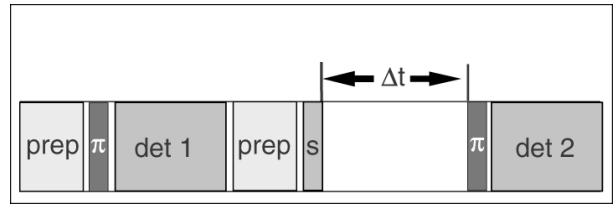}
\caption{Simplified pulse scheme for the D$_{3/2}$ lifetime measurement. It
consists of a measurement of the $\pi$-pulse transfer efficiency on the
S$_{1/2}$-D$_{5/2}$ transition (prep, $\pi$ and det1); D$_{3/2}$-state
preparation (prep, s); waiting period $\Delta t$ and state detection ($\pi$ and
det2). For details of the pulse sequence see text. The waiting time is varied
between 25~ms and 5~s.}\label{D32lt-pulse-simple}
\end{center}
\end{figure}

For the measurement of the D$_{3/2}$-level lifetime some alterations in the
pulse sequence are required (see Fig.~\ref{D32lt-pulse-simple}). To populate
the D$_{3/2}$ state we use indirect shelving by driving the S-P transition at
397~nm and taking advantage of the 1:16 branching ratio into the D$_{3/2}$
level. After a few microseconds the D$_{3/2}$ level is populated with unity
probability.

Furthermore, because that level is part of the closed 3-level fluorescence
cycle used for state detection its population cannot be probed with the state
detection scheme described in the previous paragraph. In that sense the
D$_{3/2}$ level is not a shelved state. The method used here is that prior to
state detection the decayed population is transferred to the D$_{5/2}$ shelving
state. The measured excitation probability of the D$_{5/2}$ state divided by
the shelving probability then corresponds to the decay probability from the
D$_{3/2}$ level and the further analysis is analogous to the one in
Sec.~\ref{secD52lifetime}. Shelving in the D$_{5/2}$ state is achieved by
coherent excitation ($\pi$-pulse). However, it must be taken into account that
the D$_{3/2}$ state may decay into both Zeeman sublevels of the S$_{1/2}$
ground state. Hence, two $\pi$~pulses from both sublevels are required to
transfer all decayed population to the D$_{5/2}$-state. In our experiment we
chose the two $\Delta m_j=2$ transitions ($m_j = -1/2$ to $m_j=-5/2$ and
$m_j=1/2$ to $m_j = 5/2$). The combined transfer efficiency $P_{\pi}$ of the
two pulses is determined in the first part of the pulse sequence (cf.
Fig.~\ref{D32lt-pulse-simple}): after Doppler cooling the ion is not optically
pumped into the S$_{1/2} (m_j=-1/2)$ ground state as usual but might populate
both Zeeman sublevels. The measured transfer efficiency $P_{\pi}$ is used for
calculation of the decay probability.

The measurement of the D$_{3/2}$-level lifetime $\tau_{\textrm{(3/2)}}$ thus
consists of a repetition of the following laser pulse sequences applied to the
ion. The sequence generally is composed of three steps (cf.
Fig.~\ref{D32lt-pulse-simple}):

1. Measurement of transfer efficiency $P_{\pi}$: state preparation and Doppler
cooling, consisting of 2~ms of Doppler cooling (397~nm and 866~nm light),
repumping from the D$_{5/2}$ level (854~nm light) and spontaneous decay into
the S$_{1/2} ($m=-1/2$)$ or $($m=+1/2$)$ Zeeman sublevel; $\pi$-pulses on the
S$_{1/2}$ to D$_{5/2}$ transitions ($m_j = -1/2$ to $m_j=-5/2$ and $m_j=1/2$ to
$m_j = 5/2$); state detection for 3.5 ms by recording the fluorescence on the
S$_{1/2}$ - P$_{1/2}$ transition with a photomultiplier.

2. State preparation and shelving in the D$_{3/2}$-level: 2~ms of Doppler
cooling (397~nm and 866~nm light), repumping from the D$_{5/2}$ level (854~nm
light) and optical pumping into the S$_{1/2} ($m=-1/2$)$ Zeeman sublevel
(397~nm $\sigma^-$ polarized light); shelving pulse at 397~nm for a few $\mu$s.

3. Measurement of decay probability: free decay for a variable delay time;
$\pi$-pulses on the S$_{1/2}$ to D$_{5/2}$ transitions ($m_j = -1/2$ to
$m_j=-5/2$ and $m_j=1/2$ to $m_j = 5/2$); state detection for 3.5 ms by
recording the fluorescence on the S$_{1/2}$ - P$_{1/2}$ transition with a
photomultiplier.

\begin{figure}[!htb]
\includegraphics[width=8cm]{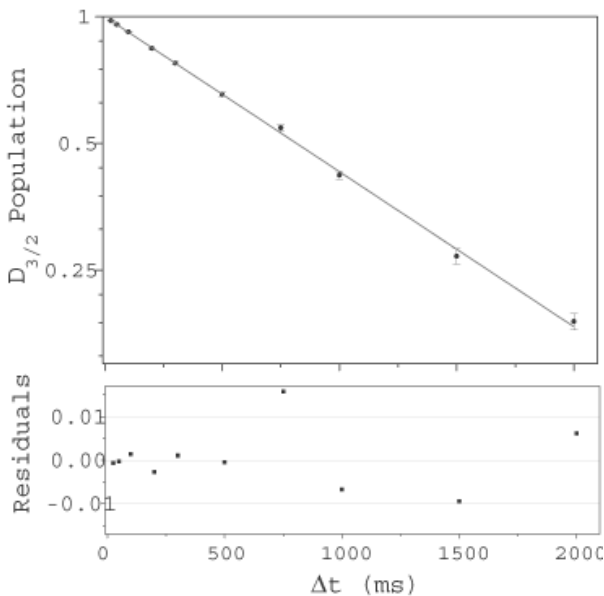}\caption{$D_{3/2}$-level
excitation probability for delay times from 25~ms to 2~s plotted on a
logarithmic scale. The solid line is a least squares fit to the data yielding
$\tau_{\textrm{(3/2)}}$ = 1176(11)~ms. The residuals (difference of data points
and fit curve) of the fit are shown in the lower diagram.}\label{D32lifetime}
\end{figure}

The measured D$_{3/2}$-level excitation probability $(1-p)$ is plotted vs.
various delay times between 25~ms and 2~s in Fig.~\ref{D32lifetime}. Again, the
data have been fitted using the least squares method and the fit function $1-p
= \exp(-\Delta t/\tau_{\textrm{(3/2)}})$. Here, $p$ denotes the corrected decay
probability $p=P_{ex}/P_{\pi}$, i.e. the detected excitation of the D$_{5/2}$
level $P_{ex}$ corrected for the near-unity shelving probability $P_{\pi}$
(which is typically 0.98-0.99 on average). Since there is no correlation
between $P_{\pi}$ and $P_{ex}$ in one experiment it is more appropriate to use
for the correction the average of $P_{\pi}$ for each delay time. The output
from the fit is $\tau_{\textrm{(3/2)}}$ = 1176(11)~ms with $\chi^2_{\nu}=0.68$
indicating good agreement of data and exponential model.


\subsection{Systematic errors}

Also for this experiment systematic errors due to residual light have to be
investigated. The measured lifetime might be reduced by residual light at
866~nm or 850~nm present during the delay interval. This light would de-excite
the D$_{3/2}$ level via the P$_{1/2}$- or P$_{3/2}$-levels, respectively, and
results in a faster effective decay rate. The main source of light at 866~nm is
the corresponding diode laser itself which is switched with a single pass AOM
with an attenuation of 20~dB. As this attenuation was found to be insufficient
a fast mechanical shutter (cf. Sec.~\ref{secD52lifetime}) was installed which
remained closed throughout the entire waiting period. The fluorescence
background of the 854~nm diode laser at 866~nm was found to be negligible.
Light at 850~nm could de-shelve the ion via the P$_{3/2}$ state and is expected
to mainly originate from the fluorescence background of the 854~nm diode laser.
For this laser, a double-pass AOM attenuation of 40~dB was proven to be
sufficient since no effect on the lifetime could be measured within a 5\% error
even if the laser was switched on at full intensity during the whole waiting
time.

Lifetime reducing effects are not obviously detectable because they only
increase the decay rate while the functional shape of the decay curve remains
the same, i.e. no offset is introduced for long delay times. The main concern
in our experiment is the 866~nm light and extreme care has been taken to ensure
that the shutter was indeed closed during the delay time. Before the 397~nm
shelving pulse and between the $\pi$~pulse and the second detection a 1~ms
period has been inserted to allow for shutting time and jitter. During the
lifetime measurements the correct shutting was checked by monitoring the
transmission on photodiodes. In fact the shutters close fast within about
400~$\mu$s but the start time is not well defined and jitters by about
500~$\mu$s.

Lifetime prolonging effects can be induced by residual light at 397~nm or
729~nm which might re-excite the ion after it has already decayed. This
re-shelving process can be detected as an offset as already pointed out in
Sec.~\ref{secD52lifetime}. The 397~nm laser light is switched by two single
pass AOM´s in series (one before a fiber, one behind it, combined attenuation
$\approx$ 55~dB). Nonetheless, we used a shutter to exclude the influence of
397~nm laser light to the largest possible extent. To give a limit on the
systematic effect of any re-pumping source the same method as in section
\ref{syserr52} is applied. The experimental data is fitted with the rate model
function Eq.~(\ref{pD}) yielding a rate of $R_{\textrm{fit}} = 3(10) \times
10^{-3}$~s$^{-1}$. The standard deviation of $R$ for an simulated ideal data
set is $\Delta R = 1.5 \times 10^{-2}$~s$^{-1}$ (with mean $R=0$), so again the
rate is concealed by the statistical error. From simulations an upper limit for
the systematic lifetime error of $\Delta \tau= -2$~ms is obtained.

\begin{figure}[!htb]
\includegraphics[width=8cm]{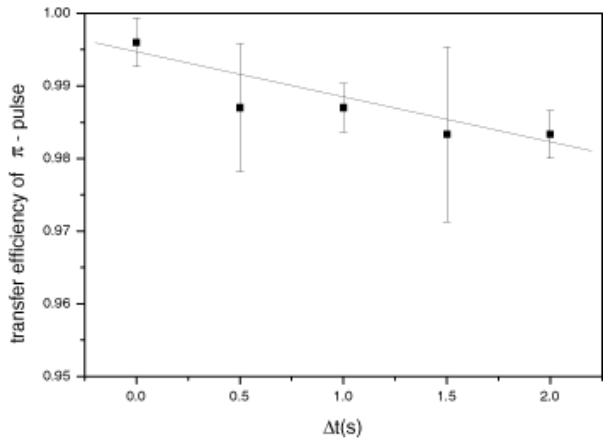}
\caption{Average transfer efficiency  of the $\pi$-pulses on the S$_{1/2}$ to
D$_{5/2}$ transition after various delay times between Doppler cooling and
$\pi$-pulses.}\label{pipulse-eff}
\end{figure}

Another source of systematic error is vibrational heating of the ion during the
delay time. If, due to heating, the transfer efficiency $P_{\pi}(\Delta t)$ is
smaller after the waiting time than $P_{\pi}(0)$ determined in the first part
of the pulse sequence the correction for the transfer efficiency is too small
and the actual decay rate is higher than measured. A $\pi$~pulse only has high
transfer efficiency if the ion is in the Lamb-Dicke regime
\cite{Leibfried03,SasuraReview}, i.e. $\eta^2\bar{n}<<1$ where $\eta$ is the
Lamb-Dicke parameter and $\bar{n}$ is
the mean phonon number. 
If the factor $\eta^2\bar{n}$ becomes significant both the Rabi frequency
$\Omega_{\bar{n}}$ and the maximum transfer efficiency decrease as
$\Omega_{\bar{n}} = \Omega_0 (1-\eta^2\bar{n})$, where $\Omega_0$ is the
coupling strength on the S-D transition. 
Taking the mean phonon number after Doppler cooling of $\bar{n} \approx 10$ and
the measured heating rate in the linear ion trap of $\partial n / \partial t
\approx 10$~s$^{-1}$ \cite{Roos99} we can estimate the transfer efficiency
after a waiting time of 2~s, $P_{\pi}(2~\textrm{s})=0.98$ if the $\pi$-pulse
time was initially chosen to fulfill $P_{\pi}(0)=1$. We experimentally checked
the degradation of transfer efficiency with waiting time by introducing a delay
time $\Delta t$ between the Doppler cooling pulse and the $\pi$-pulse in the
first step of the pulse sequence and subsequently performing a state detection
measurement. Figure~\ref{pipulse-eff} shows an average of various measurements
of $\pi$-pulse transfer efficiency $P_{\pi}(\Delta t)$ vs. delay time $\Delta
t$.
A linear fit $P_{\pi}(\Delta t)=1 - a \Delta t$ to the data yields $a=-4(2)
\times 10^{-3}$~s$^{-1}$. With simulated data sets including such a decreasing
transfer efficiency $P_{\pi}(\Delta t)$ we determine a systematic lifetime
error of $\Delta \tau= -7$~ms.

Finally, the detection error is considered, in analogy to section
\ref{syserr52}. From simulated data sets with a detection error of
$\varepsilon_2 = 1\times 10^{-3}$ a systematic error for the lifetime of
$\Delta \tau = +8 $~ms is found. Systematic errors due to collisional effects
(quenching and j-mixing) can be again neglected as argued above.

Summarizing the analysis, the lifetime for the $D_{3/2}$ level is given as:
$\tau_{\textrm{(3/2)}}=1176$~ms $\pm 11$~ms~(statistical) -2~ms~(repumping
rate) -7~ms~(heating) +8~ms~(detection error).


\section{Ab initio all order calculation of the D-states
lifetimes}

We conducted the calculation of the 
3d $^2$D$_{3/2}$ - 4s $^2$S$_{1/2}$ and 3d $^2$D$_{5/2}$ - 4s $^2$S$_{1/2}$
electric-quadrupole matrix elements in Ca$^+$ using a relativistic all-order
method which sums infinite sets of many-body perturbation theory terms. These
matrix elements are used to evaluate the $3d$-level radiative lifetimes and
their ratio.

In this particular implementation of the all-order method, the wave function of
the valence electron $v$ is represented as an expansion
\begin{eqnarray}
 |\Psi_v \rangle &= &\left[ 1 + \sum_{ma} \, \rho_{ma}
a^\dagger_m a_a + \frac{1}{2} \sum_{mnab} \rho_{mnab} a^\dagger_m a^\dagger_n
a_b a_a +
 \right. \nonumber \\
&+& \left. \sum_{m \neq v} \rho_{mv} a^\dagger_m a_v + \sum_{mna}
\rho_{mnva} a^\dagger_m a^\dagger_n a_a a_v \right. \nonumber \\
&+&  \left.\frac{1} {6}\sum_{mnrab} \rho_{mnrvab}
 a_m^{\dagger } a_n^{\dagger} a_r^{\dagger } a_b a_aa_v \right]|
\Phi_v\rangle  , \label{eq1}
\end{eqnarray}
where $\Phi_v$ is the lowest-order atomic wave function, which is taken to be
the {\em frozen-core} Dirac-Hartree-Fock (DHF) wave function of a state $v$.
This lowest-order atomic wave function can be written as $ |\Phi_v\rangle
=a_v^{\dagger }|0_C\rangle, $ where $|0_C\rangle $ represents DHF  wave
function of a closed core. In equation (\ref{eq1}), $a^\dagger_i$ and $a_i$ are
creation and annihilation operators, respectively. The indices $m$, $n$, and
$r$ designate excited states and indices $a$ and $b$ designate core states. The
first two lines of Eq.~(\ref{eq1}) represent the single and double excitation
terms. The restriction of the wave function to the first five terms of
Eq.~(\ref{eq1}) represents a single-double (SD) approximation. The last term of
Eq.~(\ref{eq1}) represents a class of the triple excitations and is included in
the calculation partially as described in Ref.~\cite{relsd}. We carried out the
all-order calculation with and without the partial addition of the triple term;
the results of those two calculations are labeled SD (single-double)  and SDpT
(single-double partial triple) data in the text and tables below. The
excitation coefficients $\rho_{ma}$, $\rho_{mv}$, $\rho_{mnab}$, and
$\rho_{mnva}$ are obtained as the iterative solutions of the all-order
equations in the finite basis set. The basis set, used in the present
calculation, consists of the single-particle states, which  are linear
combinations of B-splines \cite{bsplines}. The single-particle orbitals are
defined on a non-linear grid and are constrained to a spherical cavity.
The cavity radius is chosen to accommodate the $4s$ and $3d$ orbitals.

The matrix element of the operator $Z$ for the transition between the states
$w$ and $v$ is obtained from the expansion (\ref{eq1}) using
 \begin{equation}
Z_{wv}=\frac{\langle \Psi_w |Z| \Psi_v \rangle}{\sqrt{\langle \Psi_v | \Psi_v
\rangle \langle \Psi_w | \Psi_w \rangle}}. \label{eqr}
\end{equation}
The resulting expression for the numerator of the Eq.~(\ref{eqr}) consists of
terms that are linear or quadratic functions of the excitation coefficients. We
refer the reader to Refs.~\cite{sd,cs,relsd} for further description of the
all-order method.

\begin{table} [ht]
\caption{\label{tab1} Electric-quadrupole reduced matrix elements $E2$ in
Ca$^+$ (in a.u.)
 calculated using different
approximations: Dirac-Hartree-Fock (DHF), third-order many-body perturbation
theory (Third order), single-double all-order method (SD), and single-double
all-order method including  partial triple excitation contributions (SDpT). The
all-order data calculated with $l_{\max}=6$ are listed separately. The
contribution of basis set states with orbital angular momentum $l=7-10$
calculated using third-order MBPT is listed in rows labeled ``Extr.''. This
correction is added to obtain the values listed in rows labeled ``SD'' and
``SDpT''. }
 \begin{ruledtabular}
\begin{tabular}{lrr}
\multicolumn{1}{c}{Transition}& \multicolumn{1}{c}{Method}&
\multicolumn{1}{c}{Value} \\
\hline
3d $^2$D$_{3/2}$ - 4s $^2$S$_{1/2}$  & DHF          &  9.767        \\
               & Third order  &   7.364            \\
               & SD$_{l_{max}=6}$   &  7.788      \\
               & SDpT$_{l_{max}=6}$ &  7.971   \\
               & Extr.\footnotemark[1]&  -0.038   \\[0.1cm]
               & SD           &  7.750   \\
               & SDpT         &  7.934    \\    [0.2cm]
               \hline
3d $^2$D$_{5/2}$ - 4s $^2$S$_{1/2}$  & DHF          &  11.978   \\
               & Third order  &   9.046            \\
               &   SD$_{l_{max}=6}$&  9.562     \\
               & SDpT$_{l_{max}=6}$&  9.786   \\
               & Extr.\footnotemark[1]&  -0.046   \\[0.1cm]
               & SD           &  9.516 \\
               & SDpT         &  9.740 \\
\noindent \footnotetext[1]{This value is the difference of the third-order
result obtained with the same basis set as the all-order calculation (the
number of splines $N=35/40$ and $l_{max}=6$) and
third-order result with $N=70$ and $l_{max}=10$.} \\
\end{tabular}
\end{ruledtabular}
\end{table}
   \begin{table*} [ht]
\caption{\label{tab2} The Breit correction to the third-order values of the 4s
$^2$S$_{1/2}$ - 3d $^2$D$_{3/2}$ and 4s $^2$S$_{1/2}$ - 3d $^2$D$_{5/2}$
electric-quadrupole matrix elements. The Dirac-Hartree-Fock values are given in
column DHF. The random-phase approximation (RPA) values, iterated to all
orders, are listed in column RPA. The third-order Brueckner-orbital, structure
radiation, and normalization terms are listed in the columns BO, SR, and Norm,
respectively.}
 \begin{ruledtabular}
\begin{tabular}{lrrrrrrr}
\multicolumn{1}{c}{Transition} & \multicolumn{1}{c}{} & \multicolumn{1}{c}{DHF}
& \multicolumn{1}{c}{RPA} & \multicolumn{1}{c}{BO} & \multicolumn{1}{c}{SR} &
\multicolumn{1}{c}{Norm} &
\multicolumn{1}{c}{Total} \\
\hline
4s $^2$S$_{1/2}$ - 3d $^2$D$_{3/2}$ & no Breit   &    9.7673  &  -0.0553 &   -2.2136 &    0.0621  &  -0.1588&     7.4018  \\
                & with Breit &    9.7611  &  -0.0552 &   -2.2131 &    0.0621  &  -0.1589&     7.3961  \\
                & Difference &   -0.0062  &   0.0001 &    0.0005 &    0.0000  &  -0.0001&    -0.0057  \\[0.2pc]
 4s $^2$S$_{1/2}$ - 3d $^2$D$_{5/2}$ & no Breit   &   11.9782  &  -0.0662 &   -2.7006  &   0.0756 &   -0.1945 &    9.0926 \\
                & with Breit &   11.9672  &  -0.0662 &   -2.7001  &   0.0756 &   -0.1946 &    9.0820 \\
                & Difference &   -0.0110  &   0.0000 &    0.0005  &   0.0000 &   -0.0001 &   -0.0106  \\
\end{tabular}
\end{ruledtabular}
\end{table*}

The numerical implementation of the all-order method requires to carry out the
sums over the entire basis set. We truncate those sums at some value of the
orbital angular momentum  $l_{max}$; $l_{max}=6$ in the current all-order
calculation. The contributions of the excited states with higher values of $l$
which are small but significant for the considered transitions, are evaluated
in the third-order many-body perturbation theory (MBPT). To evaluate those
contributions, we carried out a third-order MBPT calculation with the same
basis set and $l_{max}$, used the all-order calculation and then repeated the
third-order calculation with larger basis set containing the orbitals with $l$
up to $l_{max}=10$. The difference between these two results is added to the
{\it ab initio} all-order results. The convergence of the MBPT terms with $l$
is rather rapid; the differences between the third-order calculations with
$l_{max}=4, 6, 8, 10$ are 1.8\%, 0.4\%, and 0.1\%, respectively. The last
number is well below the expected uncertainty of the current calculation. Thus,
the contribution from orbitals with  $l_{max}>10$ can be omitted at the present
level of accuracy. The contribution from the excited states with $l_{max}>6$
relative to the total value of the matrix elements is significantly larger for
$4s-3d$ electric-quadrupole transitions (about 0.5\%) than for the primary
$ns-np$ electric-dipole transitions in alkali-metal atoms. We note that while
the all-order matrix elements contain the entire third-order perturbation
theory contribution there is no straightforward and simple way to directly
separate it out (see Ref.~\cite{SafronovaPhD} for the all-order vs.
perturbation theory term correspondence issue). Thus, we have conducted a
separate third-order calculation following Ref.~\cite{ADNDT}. The results of
the third-order and the all-order calculations (with and without partial
inclusion of the triple excitations) are listed in Table~\ref{tab1}. The
contribution from the excited states with orbital angular momentum $l>6$
calculated as described above is listed in the row labeled ``Extr.''. The
all-order values corrected for the truncation of the higher partial waves are
listed in rows labeled ``SD'' and ``SDpT''.

We also investigated the effect of the Breit interaction to the values of the
electric-quadrupole matrix elements. The Breit interaction arises from the
exchange of a virtual photon between atomic electrons. The static Breit
interaction can be described by the operator
 \begin{equation}
 B_{ij}= - \frac {1}{r_{ij}} \hspace{1mm}
\mbox{\boldmath $\alpha$}_i \cdot \mbox{\boldmath $\alpha$}_j +
\frac{1}{2r_{ij}} \left[\hspace{1mm} \mbox{\boldmath $\alpha$}_i \cdot
\mbox{\boldmath $\alpha$}_j - \left( \mbox{\boldmath $\alpha$}_i \cdot
\mathbf{\hat{r}}_{ij} \right) \left(\mbox{\boldmath $\alpha$}_j \cdot
\mathbf{\hat{r}}_{ij} \right)\right] \label{breiteq}
\end{equation}
where the first part results from instantaneous magnetic interaction between
Dirac currents and the second part is the retardation correction to the
electric interaction \cite{shells}. In Eq.~(\ref{breiteq}), \mbox{\boldmath
$\alpha$}$_i$ are Dirac matrices. The complete expression for the Breit matrix
elements is given in \cite{breit}. To calculate the correction to the matrix
elements arising from the Breit interaction, we modified the generation of the
B-spline basis set to intrinsically include the Breit interaction on the same
footing as the Coulomb interaction and repeated the third-order calculation
with the modified basis set. The difference between the new values and the
original third-order calculation (conducted with otherwise identical basis set
parameters) is taken to be the correction due to Breit interaction. We give the
breakdown of the third-order calculation with and without inclusion of the
Breit interaction in Table \ref{tab2}. The Dirac-Hartree-Fock values are given
in column DHF. The random-phase approximation (RPA) values, iterated to all
orders, are listed in column RPA. The third-order Brueckner-orbital, structure
radiation and normalization terms are listed in the columns BO, SR, and Norm,
respectively. The breakdown of the third-order calculation to RPA, BO,
structure radiation and normalization terms follows that of Ref.~\cite{ADNDT}.
The reader is referred to Ref.~\cite{ADNDT} for the detailed description of the
third-order MBPT method and the formulas for all of the terms. We find the
Breit correction to the DHF contribution to be dominant, with the contributions
from all other terms being insignificant. The total Breit correction is very
small and below the estimated uncertainty of our theoretical values discussed
below. However, the Breit contribution to the ratio of the matrix elements is
found to be small but significant owing to higher accuracy of the ratio.

The procedure described above does not include a class of the Breit correction
contributions referred to in Ref.~\cite{andrei_breit} as two-body Breit
contribution \cite{Breitfootnote}. To conduct the study of the  possible size
of the two-body Breit contribution we calculated the Breit contribution to 10
different electric-dipole matrix elements ($6s-6p$, $6s-7p$, $7s-7p$, $7s-6p$,
and $5d_{3/2}-6p$) in Cs using the method described above and compared those
values with the results from \cite{andrei_breit}. Cs is chosen as ``model''
atom as it is a similar system compared to Ca$^+$. The Breit contribution to Cs
properties was studied in detail owing to its importance for the interpretation
of Cs parity nonconservation experiments. In Ref.~\cite{andrei_breit}, both
one-body and dominant two-body Breit contributions have been taken into
account. We find the largest difference between our data and that of
\cite{andrei_breit} to be 25\%. For most of the transitions, we either agree to
all the digits quoted in \cite{andrei_breit} or differ by 10\% or less.
Therefore, the two-body contribution was not significant for any of the Cs
electric-dipole transitions that we could compare with. We agree with the
values of the Breit correction to the DHF matrix elements given in
Ref.~\cite{andrei_breit} exactly, as expected, since the two-body Breit
contribution only affects the correlation part of the calculation. Thus, we
found no evidence that the two-body Breit correction may exceed the already
calculated one-body correction, especially considering the fact that the Breit
correction to the lowest-order DHF value dominates the one-body Breit
correction to the $4s-3d$ matrix elements in Ca$^+$. Therefore, we assume that
the two-body Breit contribution does not exceed the already calculated part. In
summary, the omission of the two-body Breit interaction introduces an
additional uncertainty in our calculation and we take the uncertainty to be
equal to the value of the correction itself. Most likely, it is an overly
pessimistic assumption based on the comparison with the calculation of the
Breit correction to Cs electric-dipole matrix elements carried out in
Ref.~\cite{andrei_breit}.

Next, we use a semi-empirical scaling procedure to evaluate some classes of the
correlation correction omitted by the current all-order calculation. The
scaling procedure is described in Refs.~\cite{cs,SafronovaPhD}. Briefly, the
single-valence excitation coefficients $\rho_{mv}$ are multiplied by the ratio
of the corresponding experimental and theoretical correlation energies and the
calculation of the matrix elements is repeated with those modified excitation
coefficients. This procedure is especially suitable in this particular study
since the matrix element contribution containing the excitation coefficients
$\rho_{mv}$ overwhelmingly dominates the correlation correction for the
considered here transitions. We conduct this scaling procedure for both SD and
SDpT calculations; the scaling factors are different in these two cases as SD
method underestimates and SDpT method overestimates the correlation energy.

Table~\ref{tab3} contains the summary of the resulting matrix elements; the
Breit correction is included in all values. We note that the scaled values only
include DHF part of the Breit correction to avoid possible double counting of
the terms (because of the use of the experimental correlation energy in the
scaling procedure). The final values are taken to be scaled SD values based on
the comparisons of similar calculations in alkali-metal atoms with experiment
\cite{relsd,SafronovaPhD,rb}. The uncertainty is calculated as the spread of
the scaled values and {\it ab initio} SDpT values. The uncertainty in the Breit
interaction calculation is also included; it is negligible in comparison with
the spread of the values.

  \begin{table} [ht]
\caption{\label{tab3} Electric-quadrupole reduced matrix elements $E2$ in
Ca$^+$ (a.u.) }
 \begin{ruledtabular}
 \begin{tabular} {lrrr}
\multicolumn{1}{c}{Transition}& \multicolumn{1}{c}{Method}&
\multicolumn{1}{c}{{\it ab initio}} &
\multicolumn{1}{c}{scaled} \\
\hline
3d $^2$D$_{3/2}$ - 4s $^2$S$_{1/2}$  & SD           &  7.744 & 7.939   \\
               & SDpT         &  7.928 & 7.902   \\  [0.2cm]
               & Final        && 7.939(37)  \\[0.2cm]
               \hline
 3d $^2$D$_{5/2}$ - 4s $^2$S$_{1/2}$ & SD           &  9.505 &   9.740 \\
               & SDpT         &  9.729 &   9.694\\ [0.2cm]
               & Final && 9.740(47) \\[0.2cm]
\end{tabular}
\end{ruledtabular}
\end{table}


We use our final theoretical results to calculate the lifetimes of the
D$_{3/2}$ and D$_{5/2}$ states in Ca$^+$. The transition probabilities $A_{vw}$
are calculated using the formula \cite{WRJnotes}
 \begin{equation}
 A_{vw}=\frac{1.11995\times 10^{18}}{\lambda^5}
 \frac{|\langle v\|Q\|w\rangle|^2}
 {2j_v+1}\,s^{-1},
 \end{equation}
where $\langle v\|Q\|w\rangle$ is the reduced electric-quadrupole matrix
element for the transition between states $v$ and $w$ and $\lambda$ is
corresponding wavelength in \AA. The lifetime of the state $v$ is calculated as
   \begin{equation}
\tau_v=\frac{1}{\sum_{w}A_{vw}}. \label{eq3}
 \end{equation}
In both D$_{3/2}$ and D$_{5/2}$ lifetime calculations we consider a single
transition contributing to each of the lifetimes. The transition probabilities
of other transitions (M1 D$_{3/2}$ - S$_{1/2}$, M1 D$_{5/2}$ - D$_{3/2}$, and
E2 D$_{5/2}$ - D$_{3/2}$) have been estimated in Ref.~\cite{Ali88} and have
been found to be 6 to 13 orders of magnitude smaller that the transition
probabilities of the D$_{3/2}$ - S$_{1/2}$ and D$_{5/2}$ - S$_{1/2}$ E2
transitions. Thus, we neglect these transitions in the present study. The
experimental energy levels from Ref.~\cite{NIST} are used in our calculation of
the lifetimes. From the calculations we yield $\tau_{(3/2)} = 1196(11)$~ms for
the D$_{3/2}$-state and $\tau_{(5/2)} = 1165(11)$~ms for the D$_{5/2}$-state.
These lifetime values are compared with experimental and other theoretical
results in Figs.~\ref{D52allresults} and \ref{D32allresults}.

 \begin{table*} [ht]
\caption{\label{tab5} The ratio of the D$_{3/2}$ and D$_{5/2}$ lifetimes in
Ca$^{+}$ in various approximations. The lowest-order Dirac-Hartree-Fock results
are labelled ``DHF'', third-order many-body perturbation theory results are in
column labelled ``Third'', the results of the {\it ab initio} all-order
calculation including single and double excitations are labeled  ``SD'', the
results of the {\it ab initio} all-order calculation including single and
double excitations with partial  addition of the triple excitations are labeled
``SDpT'', and the results of the corresponding scaled calculations are given in
columns labeled ``SD$_{sc}$'' and ``SDpT$_{sc}$'', respectively.}
 \begin{ruledtabular}
\begin{tabular}{lccccccc}
\multicolumn{1}{c}{}& \multicolumn{1}{c}{DHF}& \multicolumn{1}{c}{Third}&
\multicolumn{1}{c}{SD}& \multicolumn{1}{c}{SDpT}&
\multicolumn{1}{c}{SD$_{sc}$}& \multicolumn{1}{c}{SDpT$_{sc}$} &
\multicolumn{1}{c}{Final} \\
\hline
 No Breit     &     1.0251  &  1.0286 &   1.0275  &  1.0272 &   1.0266  &  1.0267&\\
With Breit    &     1.0245  &  1.0278 &   1.0267  &  1.0265 &   1.0259  &  1.0260 & 1.0259(9) \\
\end{tabular}
\end{ruledtabular}
\end{table*}

The all-order calculation is in agreement with the present experimental values
and recent experiments \cite{Barton00,KnoopArxiv,Staanum04} within the
uncertainty bounds. The present calculation includes the correlation
correction, which is large (23\%) for the considered transitions, in the most
complete way with comparison to all other previous calculations
\cite{Ali88,Zeippen90,Guet91,Vaeck92,Liaw95} and is expected to give the most
accurate result. It is also the only calculation which gives an estimate of the
uncertainty of the theoretical values.

In Ref.~\cite{Barton00}, the issue of the theoretical ratio of the
$\tau_{(3/2)}/\tau_{(5/2)}$ lifetimes was raised. It appeared that there was a
disagreement between previously calculated theoretical ratios; Barton {\it et
al.} \cite{Barton00} quotes the values 1.0283 \cite{Guet91}, 1.0175
\cite{Vaeck92}, and 1.0335 \cite{Liaw95}. Such a disagreement appears to be
rather puzzling since this particular ratio is far less sensitive to the
correlation correction than the values of the corresponding matrix elements.
Thus, we studied the value of the ratio and its uncertainty in detail. We list
the values of the ratio of the D$_{3/2}$ and D$_{5/2}$ lifetimes calculated in
various approximations in Table~\ref{tab5}. The experimental energy levels from
Ref.~\cite{NIST} are used in all our calculations of the lifetimes for
consistency. 
We include results with and without the addition of the Breit correction. As
mentioned before, we find that the ratio does not change substantially with the
addition of the correlation correction; in fact, the correlation only
contributes about 0.15\% to the final value. Thus, we calculate the uncertainty
in the ratio by considering the spread of the high-precision values of the
ratio itself (SD$_{sc}$, SDpT, and SDpT$_{sc}$), rather than calculating the
uncertainty in the ratio from the uncertainties in the individual matrix
elements. We also find that the while the Breit correction to the values of the
matrix elements was insignificant at the current level of accuracy this is not
the case for the ratio. In fact, the shift of the ratio values with the
addition of the the Breit interaction is of the same order of magnitude as the
spread of the high precision values as demonstrated in Table~\ref{tab5}. We
take the SD$_{sc}$ value to be our final result for consistency with the
calculation of the matrix elements. The uncertainty of the final value includes
both the uncertainty in the correlation correction contribution and the
uncertainty in the Breit interaction. As in the case of the individual matrix
elements, the uncertainty in the Breit interaction is taken to be equal to the
contribution itself. The Breit correction to the ratio is determined as the
shift in the final ratio value due to  addition of the Breit interaction.

    \begin{table} [ht]
\caption{\label{tab6} Comparison of the present values of the ratio of the
D$_{3/2}$ and D$_{5/2}$ state lifetimes in Ca$^+$ with other theory. }
 \begin{ruledtabular}
\begin{tabular}{cccccccc}
\multicolumn{1}{c}{}& \multicolumn{1}{c}{Reference}&
\multicolumn{1}{c}{Value}\\
\hline
Theory & \protect\cite{Ali88} & 1.03 \\
       & \protect\cite{Zeippen90} & 1.02 \\
       & \protect\cite{Guet91} & 1.028 \\
       & \protect\cite{Vaeck92} & 1.02 \\
       & \protect\cite{Liaw95} & 1.033 \\
       & Present          & 1.0259(9)\\
Expt.  & Present          & 1.0068(122)
\end{tabular}
\end{ruledtabular}
\end{table}

We compare our final theoretical value of the lifetime ratio with the
experiment and other theory in Table~\ref{tab6}. The ratios of the other
theoretical values \cite{Ali88,Zeippen90,Guet91,Vaeck92,Liaw95} are calculated
from the numbers in the original publications; care was taken to keep the
number of digits in the ratio consistent with the number of digits in the
values of the lifetimes or transitions rates quoted in the papers. First, we
discuss the above mentioned discrepancy of the theoretical ratios.
Ref.~\cite{Barton00} lists the following ratios: 1.0283 \cite{Guet91}, 1.0175
\cite{Vaeck92}, and 1.0335 \cite{Liaw95}. We have listed the data from the
original publications in Table~\ref{tab6} which shows that the actual numbers
with taking into account the number of digits quoted in the original papers
should have been 1271/1236=1.028 \cite{Guet91}, 1.16/1.14=1.02 \cite{Vaeck92},
and 1080/1045=1.033 \cite{Liaw95}. The first result is essentially a
third-order relativistic many-body perturbation theory calculation with
addition of semi-empirical scaling and omission of the some classes of small
but significant third-order terms. It is very close to our third-order number
1.0286 from Table~\ref{tab5}. The next paper \cite{Vaeck92} quotes only 3
digits in the lifetime values (1.16s and 1.14s) so the accuracy is insufficient
to obtain the fourth digit in the ratio. We note that the method description in
\cite{Vaeck92} is that of the non-relativistic calculation and it is unclear
how the separation to D$_{3/2}$ and D$_{5/2}$ lifetimes was made. The last
calculation yields a larger ratio but that calculation has serious numerical
issues such as taking only 20 out of 40 B-splines and including too few partial
waves. It also omits all terms except Brueckner-orbital ones and possibly even
third-order Brueckner orbital contributions, which are large. The paper is not
clear on the subject of the treatment of the higher-order contributions. Thus,
we do not consider the result of \cite{Liaw95} to be reliable. Therefore, there
are essentially no inconsistencies in the previously calculated theoretical
ratios when the accuracy of the calculations is taken into account. Our
theoretical value of the lifetime ratio is higher than the experimental value.
The spread of all values in Table~\ref{tab6}, including even lowest-order DHF
values, is so small that it does not appear probable that any omitted Coulomb
correlation or two-body Breit interaction can be responsible for the
discrepancy. The only transition which can actually reduce the value of the
theoretical ratio is the D$_{3/2}$-S$_{1/2}$ M1 transition. Thus, an accurate
calculation of this transition rate will be useful in search for a theoretical
explanation of the discrepancy.  However, the transition rate published in
\cite{Ali88} is extremely small ($A_{M1}=7.39\times 10^{-11} s^{-1}$) and has
to be incorrect by many orders of magnitude to affect the ratio at such a level
which does not appear likely since the same calculation gives a reasonably good
(within 18\%) number for the D$_{3/2}$-S$_{1/2}$ E2 transition rate.


\section{Discussion}

Figures \ref{D52allresults} and \ref{D32allresults} show an
overview of the most recent experimental and theoretical results
for the lifetime of the D$_{5/2}$ and D$_{3/2}$ states, respectively, in an
chronological order. It is remarkable that the theoretical predictions scatter
rather widely, with no noticeable convergence while the experimental results
show a trend towards longer lifetimes in the recent years as more systematic
errors are identified and stamped out.

In comparison with previous work it can be concluded here that our lifetime
result for the D$_{5/2}$ level agrees with and thereby confirms the most
precise value of Barton \textit{et al.}. We stress that this lifetime
measurement is an independent check of earlier results as we used a different
measurement technique. In addition, the result for the D$_{3/2}$ level
represents the first single ion measurement and reduces the statistical
uncertainty of the previous values for the lifetime by a factor of four.

For the calculated lifetimes we find excellent agreement of the theoretical
all-order lifetimes with the experimental results. Such agreement demonstrates
the necessity of including partially the triple contributions to the all-order
calculations for these types of transitions and confirms that scaling of the
single-double all-order results, which is significantly simpler and less time
consuming calculation in comparison with {\it ab initio} inclusion of partial
triple excitations, is adequate for these types of transitions. This is an
important result for the evaluation of the accuracy of similar theoretical
calculation in Ba$^+$ which is important to parity violating experiments in
heavy atoms. Such experiments are aimed at the tests of the Standard model of
the electroweak interaction  and at the study of the nuclear anapole moments.
One of the features of most PNC studies in heavy atoms is the need for
comparable accuracy
of theoretical and experimental data.  
The current study is also of interest in regard to recently found discrepancy
between the $5d$ lifetimes and the $6s-6p$ Stark shifts in Cs \cite{uscs}.
Atomic properties of cesium were studied extensively by both experimentalists
and theorists owing to a high-precision measurement of parity non-conserving
amplitude in this atom. Both of these quantities depend on the values of the
$5d-6p$ matrix elements. While those matrix elements are the electric-dipole
ones rather than the electric-quadrupole ones studied here, the calculation
itself as well as the breakdown of the correlation correction terms is very
similar to the present calculation. Thus, the current study presents an
important benchmark in the field of high-precision measurements and
calculations. The study of the lifetime ratio demonstrated that the Breit
interaction, which produces only a very small correction to the values of the
actual matrix elements, is important in high-precision calculations of the
corresponding matrix element ratios.

\section{Acknowledgements}

This work is supported by the Austrian 'Fonds zur F\"orderung der
wissenschaftlichen Forschung' (SFB15), by the European Commission: IHP network
'QUEST' (HPRN-CT-2000-00121), Marie Curie Research Training network 'CONQUEST'
(MRTN-CT-2003-505089) and IST/FET program 'QUBITS' (IST-1999-13021), and by the
"Institut f\"ur Quanteninformation GmbH". C.~Russo acknowledges support by
Funda\c{c}\~{a}o para a Ci\^{e}ncia e a Tecnologia (Portugal) under the grant
SFRH/BD/6208/2001. H.~H. is funded by the Marie-Curie-program of the European
Union.



\begin{thebibliography}{}
\bibitem{Densitron}
Densitron, TK-CMD series. Using suitable electronics for the shutter drive,
closing and opening times of the order of 0.5~ms can be achieved. The timing
jitter is on the same time scale.
\bibitem{Breitfootnote}
The Breit interaction is a two-particle interaction just as the Coulomb
interaction. The separation of the Breit contribution to ``one-body'' and
``two-body'' parts results from transforming the Breit operator in second
quantization to normal form as described in Ref.~\cite{andrei_breit}.
\end{thebibliography}
\end{document}